\documentclass [12pt,preprint]{aastex}

\begin{document}

\title{Structure of the inner jet of OJ 287 VLBA data at 15 GHz in a super-resolution mode} 

\author{Claudio E. Tateyama\altaffilmark{1}}

\affil{CRAAM-INPE/Escola de Engenharia - Universidade Presbiteriana Mackenzie,
Rua da Consola\c c\~ao 896, 01302-907, S\~ao Paulo, SP, Brazil}

\altaffiltext{1}{CEA – - Instituto Nacional de Pesquisas Espaciais - INPE, Av. dos Astronautas, 1758, 12201-970, S\~ao Jos\'e dos Campos, SP, Brazil}

\begin{abstract}

In this work we show the results obtained from the VLBA data at 15 GHz of OJ287  in the
super-resolution mode. The data showed a jet configuration in the form of a
``fork" where superluminal components emerge via stationary components at the northwest and the southeast close to 
the core to form parallel trajectories along the southwest direction in the plane of sky. This agrees with a source structure of an extended, broad morphology of OJ287. 

\end{abstract}


\section{Introduction}

OJ287 is a well-studied BL Lac object and distinguished in the optical by its series
of prominent flares (double peaked) which recur with a period of $\sim$ 12 years \citep{sil88,vil10}.
OJ287 is the only known extragalactic source showing evidence of a major periodic component
in the optical and bringing into a scenario of a binary black hole model. 
In radio, VLBI observations of OJ287 reveal a core-jet structure  with a size not much greater than 1.5 mas. 
In 1995-96 the jet-position-angle is $\sim$ -90$^{\circ}$  which rotates in a few years to southwest and 
showing a jet position angle of $\sim$ -120$^{\circ}$ in 1998.
The change of the jet direction was modeled as a
ballistic jet precession  taking into account the optical periodicity of 12 years \citep{tak04}.    \citet{vaw12} proposed a new model rescaling the nodding model of \citet{kat97} with a precession
period of 120 years.
In 2008, OJ287 presented a dramatic change on the structure of the source.
VLBA observations at 43 GHz of \citet{agu10} showed a strong, bright knot in the southeast of the core in 2008, almost perpendicular to the previous direction of the jet.  
An analysis including earlier VLBA data at 43 GHz, \citet{agu12} found that such
transverse components were already present on the source  which was interpreted as the crossing of the jet from one side of the line of sight
to the other of the inner jet. 
In this work we investigate the structure of the inner jet of OJ287 at 15 GHz from 1995 to 2012 in the super-resolution mode \citep{tat09}.

\section{Data and Imaging procedure}

We have used the visibility VLBA data at 15 GHz from MOJAVE 
(Monitoring of jets in active galactic nuclei with VLBA experiments) program and 43 GHz from BU
(Boston University). 
Table 1 list the VLBA data used in this work. The columns of the table are: 1  
- epoch, 2 - frequency, 3  - peak brightness, 4 - synthesized beam size, and 5  - position angle of the beam. The images were obtained using the 
Difmap Package \citep{she97} and processed in an automate mode using the 
mapping script of Taylor and Shephard. The restoring beam for both data was 0.1 mas. The pixel size of the images was 0.05 mas. The major features of the maps at 43 GHz have already been presented by \citet{agu12}.

The CLEAN is a method of deconvolution invented by \citet{hog74}. The main aspects of the CLEAN can be found 
in a confront with ME (Maximun Entropy) deconvolution in e.g., \citet{ras11}. One of the aspects of the CLEAN is that it shows better images with compact features as in the case of OJ287.
The CLEAN algorithm produces a set of point components which convolved with the CLEAN beam and superposed on the residual map generates the CLEAN map. The synthesized beam (dirty beam) is obtained by averaging the outputs of all
pairs of an array. The CLEAN beam is usually a fit of the main lobe of the dirty beam a nearly Gaussian with
angular resolution $\approx  \lambda/D$ (diffraction-limited telescope). Deconvolution methods 
dealing with structure details smaller than the standard beam (diffraction pattern) are so -called the super-resolution techniques.
The interferometric data have been always subject of deconvolution techniques to obtain sharper
structures. In a weak way we can mention the super-resolution effects of MEM \citep{baj01} or the inadequacy
of the standard beam to image interferometric data \citep{rei06}. In a strong way (large factor of sharpening) we can mention the direct study of the fringe visibility by \citet{kov05} which have shown  compact
components much smaller than standard beam and the super-resolution study at 15 GHz (factor $\approx$ 5) by \citet{tat09} showing structures comparable to the structures given by the resolution at 43 GHz.
Such interferometric data along with a set of maps containing adequate  time sampling 
observations in order  to recognize features of the source in adjacent maps produce a robust data results.

The CLEAN-point image was also examined by carrying out simulation of VLBA image using 
the AIPS tasks UVCON and IMMOD. The UVCON generates visibilty data from a VLBA array geometry and the 
IMMOD modifies the
existing map image by the addition of models. The simulated data were imaged using the Difmap Package.

\section {The structure of OJ287 at 15 GHz}

Figure 1 shows the VLBA observations of OJ287 in the super-resolution mode from 1995 to 2012 at 15 GHz. 
The solid line on the maps is a reference line at position angle of -120$^{\circ}$  to guide the structure of the jet in the maps. 
We follow the identification of components adopted by Agudo et al. (2012) at 43 GHz. The bright jet feature at the eastern end of super-resolution maps denominated  ``C" in our maps 
corresponds to the core but it is not always coincident with position of the core proposed  by \citet{agu12} at 43 GHz maps. Accordingly the identification of component ``a" which is very close to the core along southeast-northwest direction also differs. In addition to the components ``C" and ``a" a discrepancy on the component ``d" was removed by introducing a feature ``d1" in our maps. Also, using the 43 GHz data, new features ``q" and ``r"  were included on the maps of 2011 and 2012.

In the 90's the source shows a narrow jet morphology dominated by two main stationary components.
One stationary component is at  $\sim$ 0.3 mas denominated ``A" and the other at $\sim$ 1 mas denominated ``G". The component  ``G" can be identified with superluminal component C6 on the VLBI Observations by \citet{tat99} before becoming a stationary component.
The others superluminal components ``K",``O", ``R",``T", ... becomes prominent knots when passing at the location
of the stationary feature at 1 mas.

It is well known that a super-resolution deconvolution is very sensitive to produce artificial structures.
While a super-resolution image of a set of compact components in a rectilinear direction is easy to capture, 
the super-resolution image of a complex structure shows complicated artifacts as arcs. An interesting aspect of such artifacts is that they contain information on the source structure 
and this can be recovered performing a simulation of VLBA observation. The simulation can iteratively fit the structure of the source (a model) with observed images (super-resolution images).
Figure 2 shows the model of the source for epochs 2005, 2009 and 2010.
The feature on the map of 2005 Jul 
showing  an arc blending with a stationary component at 0.3 mas is generated by collinear components along the jet (southeast) with a northern knot close to the core.
The transversal component in 2009 Aug is formed with a knot in the southeast of the core and
the transversal structure in 2010 May is produced by a southeast/core/northwest source configuration.

A remarkable feature on the super-resolution maps is the appearance of a strong, transverse component 
along the core in 2008-9. In fact it corresponds to the striking change on the structure of the source with
appearance of a bright knot in the southeast of the core reported by Agudo et al. (2010, 2012).
In addition to this bright transversal component in 2008 other weaker transversal features can be recognized
on the maps.
The arc (the artificial feature of the super-resolution mode) on the core of 2002 Jun is related to the knot ``L"
on the 43 GHz maps of Agudo et al. (2012), in 2003a May is related with knot ``S", in 2004 Apr-Aug-Sep is related to knot ``a" and in 2005 Jul is related
to knot ``a" in the north. These transversal features appear on the northwest or north of the core.
In 2006 Apr and 2008 May component ``a " appears on the south of core and in 2010 May the knots ``n" and ``l"  appear on the north of the core. Several knots especially early components as ``L" , ``S" or even ``a" appear as bright feature but does not show superluminal motion, suggesting a stage of carving a new channel on the jet.

In 2007 the transversal component seen at 43 GHz is not seen at 15 GHz. The component is very weak at 15 GHz
or the resolution at 15 GHz is not sufficient to produce the feature in the southeast-northwest direction. 
But in May 2008 the transversal component appears on the southeast of the core and gradually becomes very prominent
component at 15 GHz and 43 GHz.

\section {Position of the core}

The configuration of the structure at 15 GHz, in particular the transverse feature in 2008-2009, 
indicates the component ``C"  as the proper position of the core. If we choose the component in the southeast of the
core as the core it would deform the structure of the jet. 
We would have to presume that the position of the core has
changed (using the optically thin jet components as reference). The structure of the CLEAN-point maps on the 2004-2005 also indicates that the bright knots
near the core appear in the north/northwest and in the south/southeast of the core and not from the southeast to northwest as adopted by \citet{agu12}.

In this way we propose that the stationary components (bends on the flow) 
are formed in the southeast and northwest near the core.
The first knots would show bright feature near the core without producing superluminal components.   
The absence of superluminal motion would  reflect the stage 
of carving a stationary component or distorting the flux of the jet to form a broad jet. 
On this interpretation the bright knots in the southeast (south) and northwest (north) of the core
would show a form of a ``fork" with a southeast/core/northwest configuration where
superluminal components would emerge.

Figure 3 shows the 43 GHz maps at resolution of 0.1 mas. 
The solid line on the maps is the same reference line drawn on Figure 1 to guide the structure of the source
and the origin of the curve is placed on the core accordingly with SE/core/NE configuration.
The inner jet already appears in the southeast-northwest direction.
In 2007, the core is the brightest component and the transversal component is weak or very close to the
core. In 2008, the bright transversal knot is in the south of the core.

The core on the configuration of the structure of the jet around 2000 at 15 GHz  
is the brightest component at the base of the line extending from the stationary components at 0.3 mas and 1.0 mas. This configuration starts to change in 2001 with the formation of new stationary components
in the southeast-northwest direction. 
The source evolves to show parallel trajectories with superluminal components emerging 
via stationary components in the southeast and northwest close to the core (bends on the flux flow).
This agrees with a source structure of an extended, broad morphology of OJ287, 
produced by parallel trajectories than
a broad transversal expansion of a single jet.
Figure 4 shows the superposition of maps from 2008 to 2012 at 15 GHz and 43 GHz to emphasize the
``fork" aspect of the source.

The southeast/core/northwest configuration removes the abrupt variation of amplitude of core in  mid-2004 (see Figure 3 of \citealt*{agu12} ). On the scenario of a broad jet the trajectory of components is
consistent with a prominent jet in the southwest direction. 
The jet-position-angle would show large dispersion only on the inner jet. 
Furthermore the core (in the sub-parsec broad line region) would be the site of the variable gamma-ray
emission.

The striking similarity of VLBA structure at 15 GHz and 43 GHz in 2011-2012 supports the southeast/core/northwest configuration rather placing the core on the southernmost component of the transversal core region. This implies that the location of the mm core is not different of the cm core.   

\section{Conclusion}

In this work we show VLBA data at 15 GHz (MOJAVE) of  OJ287 by examining the structure in the maps in a super-resolution mode along with 43 GHz VLBA maps. The result showed that the CLEAN-point maps at 15 GHz follow
very closely the structure of the 43 GHz maps.
The innermost jet shows a configuration of SE/core/NW in the form of ``fork".  New components emerge from the core
via components at southeast and northwest near the core to show superluminal components in parallel trajectories in the southwest direction in the plane of sky. 

\acknowledgments

I am grateful to the referee for valuable suggestions and making this work a lot of easier to understand.
This research used data from MOJAVE (Lister and Homan 2005) and the 2 cm Survey (Kellermann et al. 2004) programs.
This study makes use of 43 GHz VLBA data from the Boston University gamma-ray blazar monitoring program
(http://www.bu.edu/blazars/VLBA project html) founded by NASA through the Fermi Guest Investigation Program.
The Very Long Baseline Array (VLBA) is an instrument of the National Radio Astronomy Observatory (NRAO).
NRAO is a facility of the National Science Foundation, operated by Associated Universities Inc.

\clearpage

\begin{figure}
\plotone{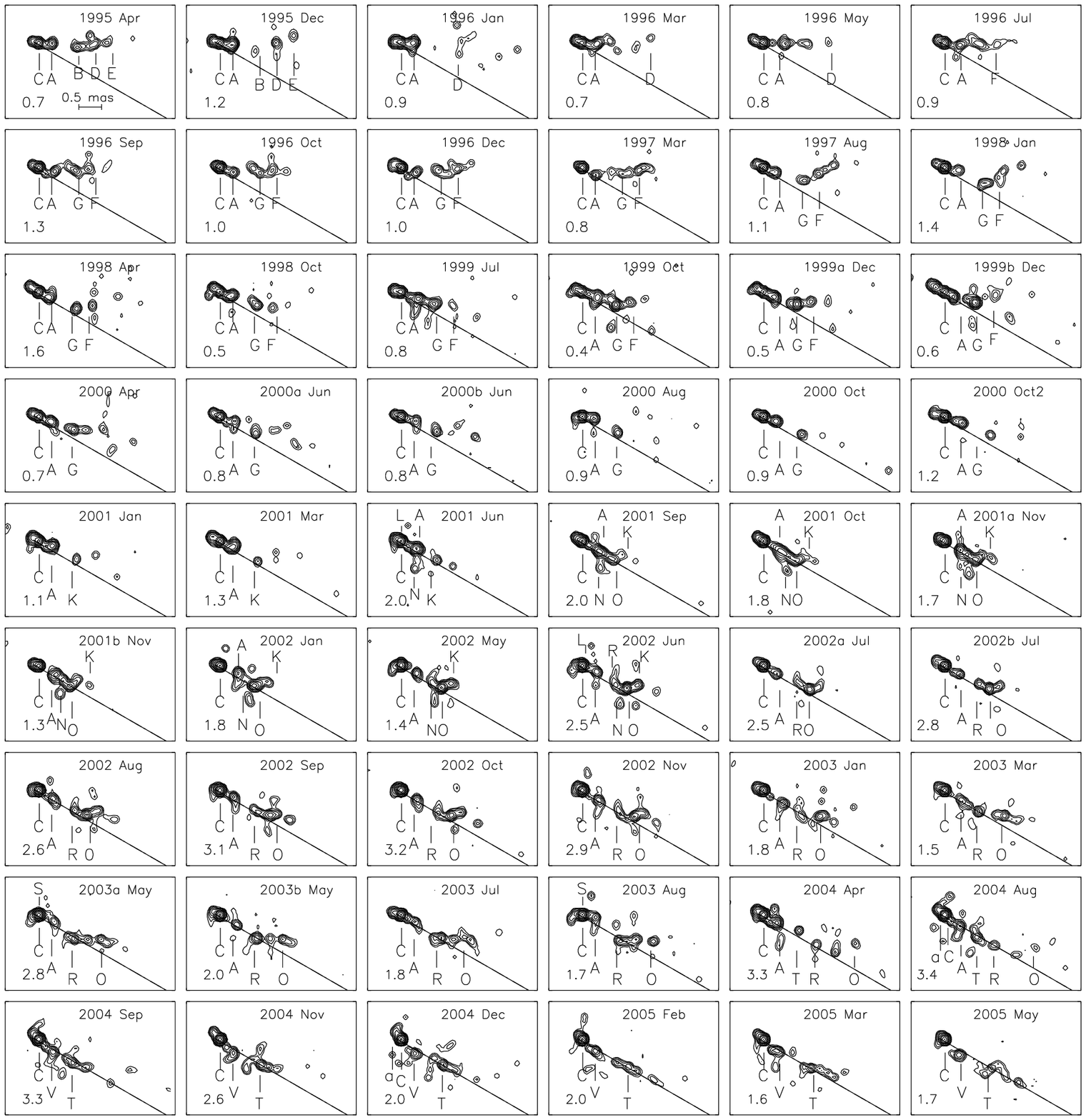}
\figcaption {15 GHz VLBA images of OJ287.  The peak flux densities of the maps are indicated 
on the left side of the maps. Contour levels are 1, 2, 4, 8, 16, 32, 64, 128, 256 and 512 $\times$ 0.003 Jy beam$^{-1}$
\label {fig1}}
\end{figure}

\addtocounter{figure}{-1}
\begin{figure}
\plotone{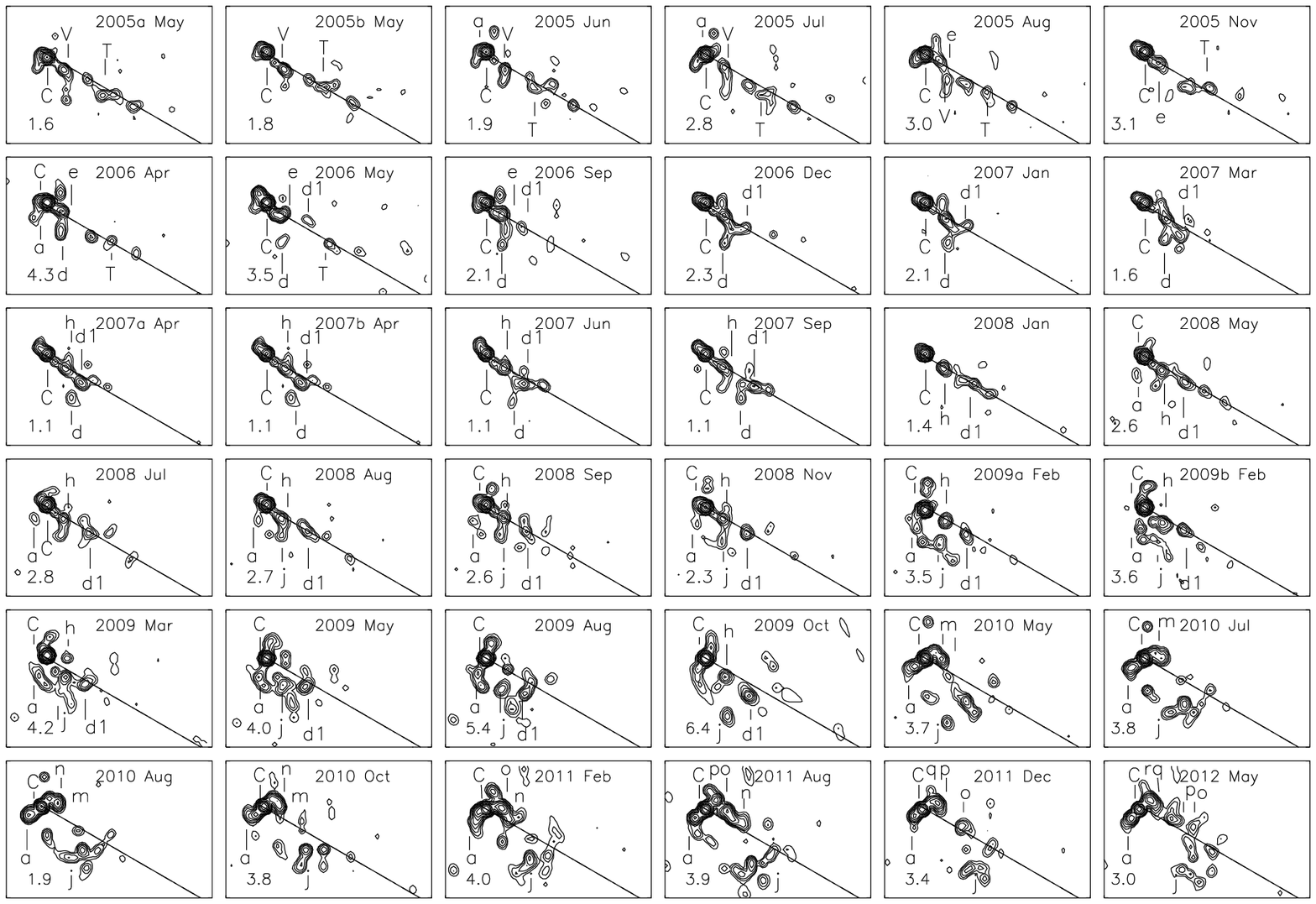}
\figcaption {(Continued)
\label{fig}}
\end{figure}

\begin{figure}
\plotone{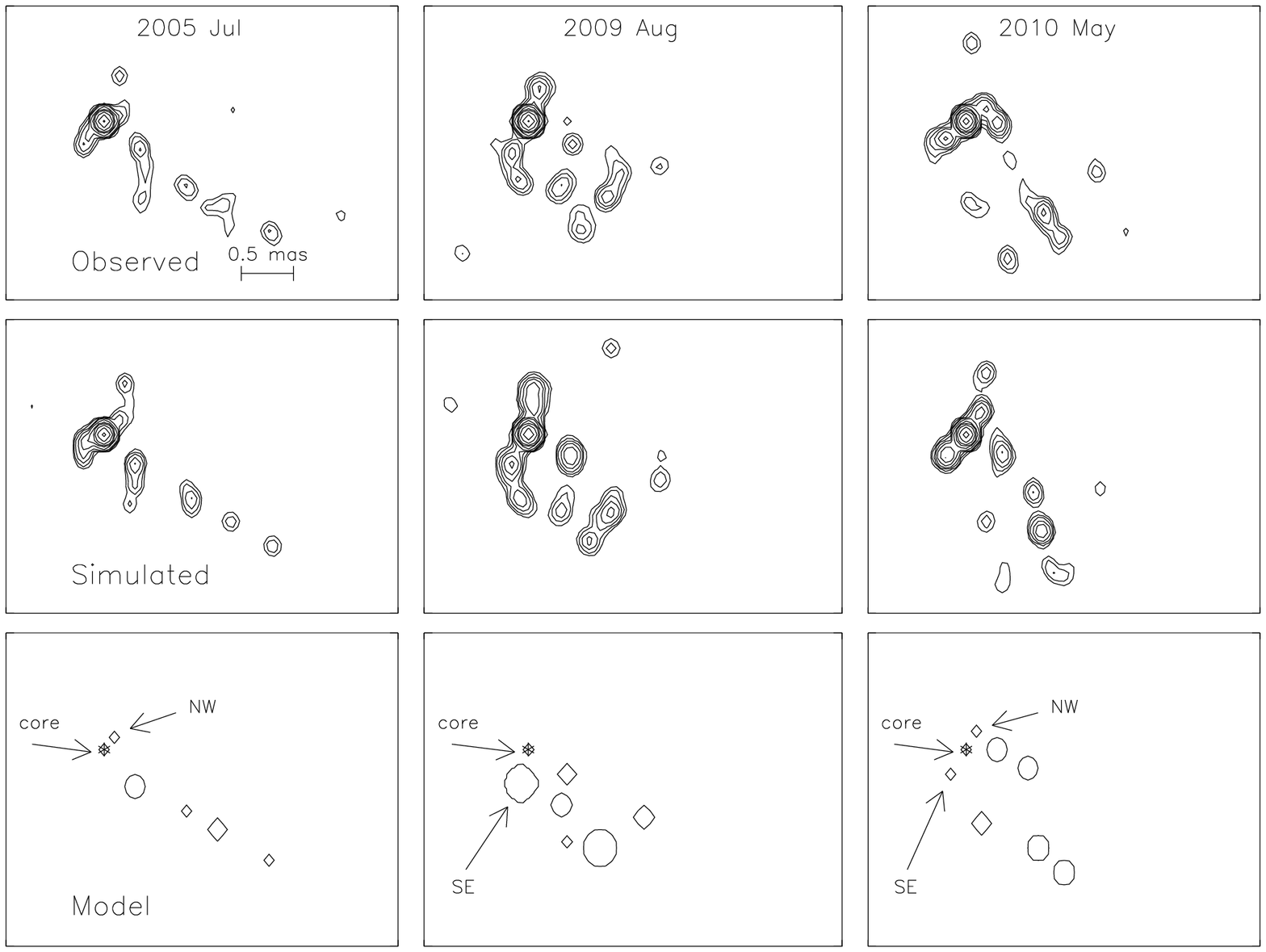}
\figcaption {Simulation and Model maps at 15 GHz.
Contour levels are 1, 2, 4, 8, 16, 32, 64, 128, 256 and 512 $\times$ 0.003 Jy beam$^{-1}$.
\label{fig2}}
\end{figure}

\begin{figure}
\plotone{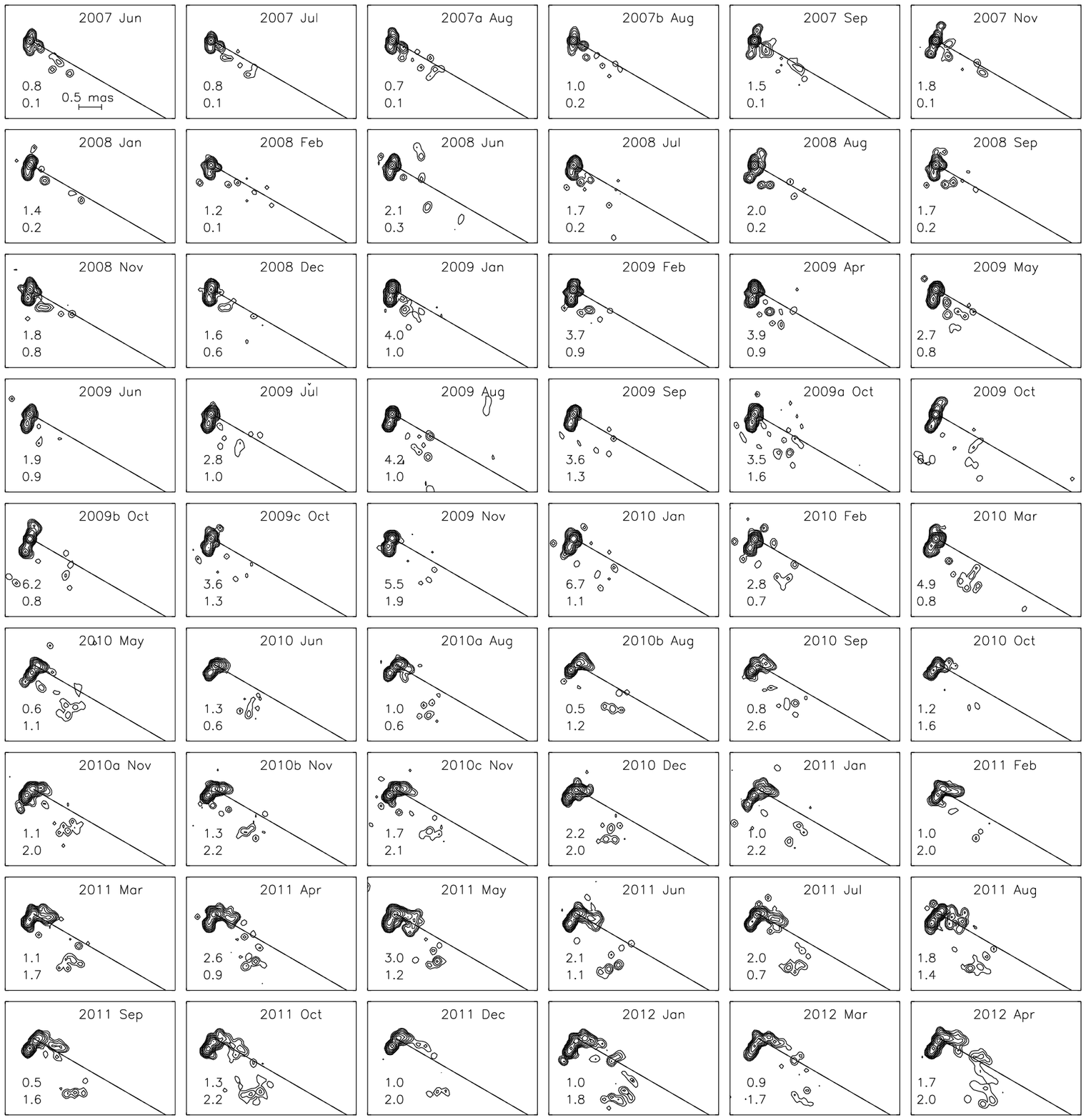}
\figcaption {43 GHz VLBA images of OJ287. The two strongest flux densities along the southeast-northwest direction are shown on the
left side of the maps, the flux on the lower position corresponds to a knot in the southeast of the core.  
Contour levels are 1, 2, 4, 8, 16, 32, 64, 128, 256 and 512 $\times$ 0.005 Jy beam$^{-1}$
\label{fig3}}
\end{figure}

\begin{figure}
\plotone{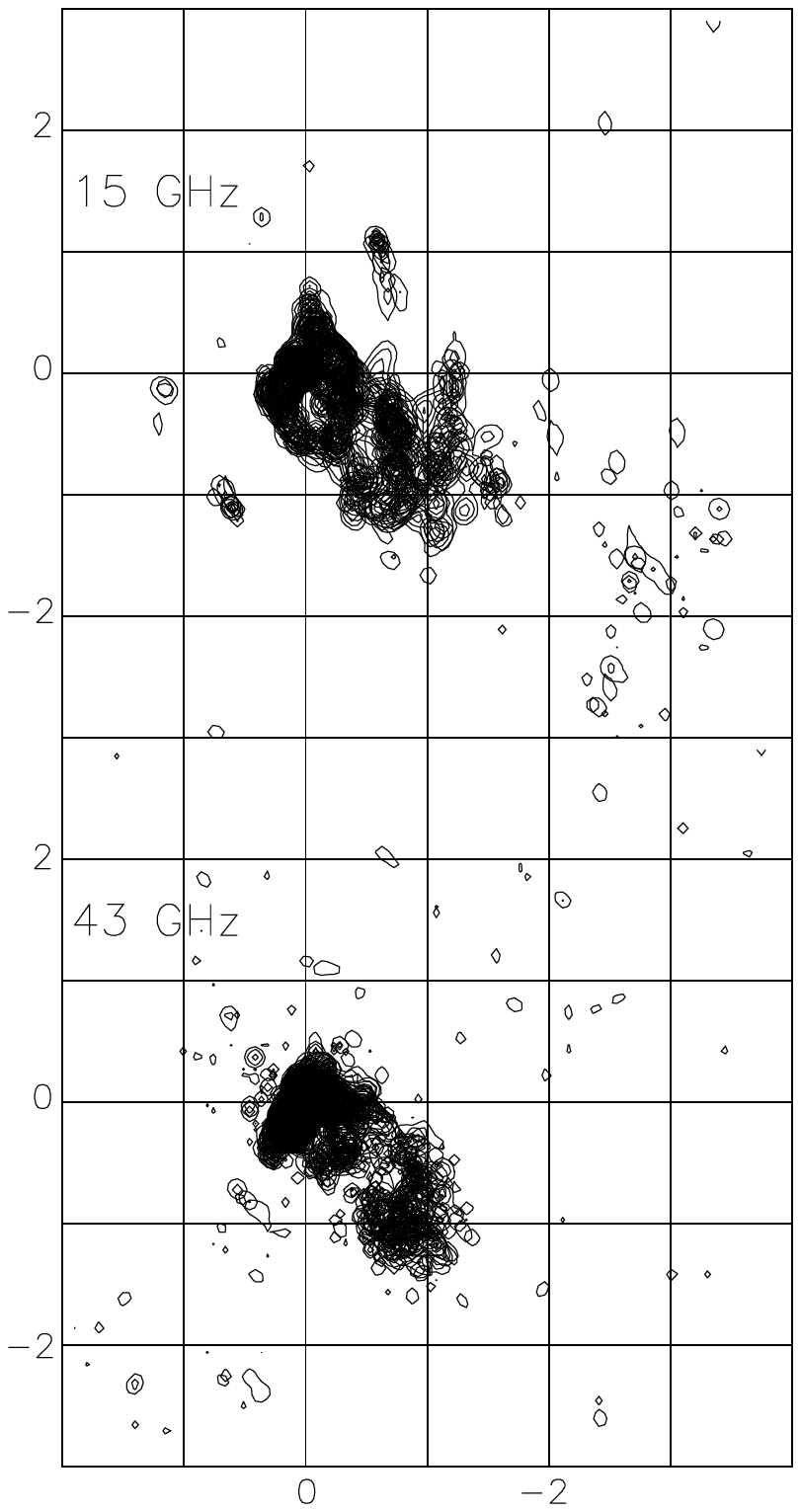}
\figcaption {Superposition of maps at 15 GHz and 43 GHz to emphasize the ``fork" aspect of the source.
Contour levels at 15 GHz are 1, 2, 4, 8, 16, 32, 64, 128, 256 and 512 $\times$ 0.0045 Jy beam$^{-1}$ and contour levels at 43 GHz are 1, 2, 4, 8, 16, 32, 64, 128, 256 and 512 $\times$ 0.005 Jy beam$^{-1}$
\label{fig4}}
\end{figure}

  \clearpage

  \begin{deluxetable} {ccccc}
  \tablecaption{VLBA data at 15 GHz (MOJAVE) and 43 GHz (BU) used in this work. \label{tbl-1}}
  \tablehead{
  \colhead{Epoch} & \colhead{Frequency} &  \colhead{I$_{peak}$} &
  \colhead{Beam} & \colhead{P.A.} \\
  & \colhead{GHz} &  \colhead{Jy beam$^{-1}$} & \colhead{mas x mas}  & \colhead{ $^{\circ}$ }
  }
  \startdata
  1995 Apr 07  & 15 & 0.9 & 1.0 x 0.5 &  7  \\
  1995 Dec 15  & 15 & 2.1 & 1.0 x 0.5 &  7  \\
  1996 Jan 19  & 15 & 1.7 & 1.0 x 0.5 &  7  \\
  1996 Mar 22  & 15 & 1.1 & 1.0 x 0.5 &  1  \\
  1996 May 27  & 15 & 1.2 & 1.1 x 0.6 & 22  \\
  1996 Jul 27  & 15 & 1.0 & 1.0 x 0.5 &  3  \\
  1996 Sep 27  & 15 & 1.4 & 1.1 x 0.5 & -1  \\
  1996 Oct 27  & 15 & 1.1 & 0.9 x 0.4 & -7  \\
  1996 Dec 06  & 15 & 1.2 & 1.2 x 0.5 & -4  \\
  1997 Mar 13  & 15 & 0.8 & 1.0 x 0.6 & -2  \\
  1997 Aug 28  & 15 & 1.3 & 1.0 x 0.5 &  3  \\ 
  1998 Jan 03  & 15 & 1.5 & 1.0 x 0.6 &  4  \\
  1998 Apr 27  & 15 & 2.1 & 1.0 x 0.6 &  3  \\
  1998 Oct 30  & 15 & 1.0 & 1.1 x 0.5 & -5  \\
  1999 Jul 24  & 15 & 1.4 & 1.2 x 0.5 & -13 \\
  1999 Oct 16  & 15 & 0.9 & 1.2 x 0.5 & -6  \\
  1999 Dec 23  & 15 & 1.1 & 1.0 x 0.5 & -1  \\
  1999 Dec 31  & 15 & 1.4 & 1.0 x 0.6 & -5  \\
  2000 Apr 07  & 15 & 1.1 & 1.1 x 0.5 & -1  \\
  2000 Jun 27  & 15 & 0.9 & 1.2 x 0.5 & -13 \\
  2000 Jun 29  & 15 & 1.0 & 1.2 x 0.5 &  1  \\
  2000 Aug  25 & 15 & 1.1 & 1.2 x 0.5 & -4  \\
  2000 Oct 08  & 15 & 1.2 & 1.1 x 0.6 &  5  \\
  2000 Oct 17  & 15 & 1.4 & 1.2 x 0.6 &  19 \\
  2001 Jan 22  & 15 & 1.9 & 1.2 x 0.5 & -11 \\
  2001 Mar 05  & 15 & 2.0 & 1.2 x 0.6 &  9  \\
  2001 Jun 30  & 15 & 2.9 & 1.1 x 0.5 &  1  \\
  2001 Sep 05  & 15 & 2.6 & 1.1 x 0.6 & -10 \\
  2001 Oct 22  & 15 & 2.2 & 1.1 x 0.6 & -8  \\
  2001 Nov 02  & 15 & 2.1 & 1.1 x 0.5 &  0  \\
  2001 Nov 07  & 15 & 1.7 & 1.1 x 0.5 & -10 \\
  2002 Jan 08  & 15 & 2.1 & 1.2 x 0.6 &  8  \\
  2002 May 01  & 15 & 1.7 & 1.1 x 0.5 & -5  \\
  2002 Jun 27  & 15 & 2.8 & 1.1 x 0.6 & -3  \\
  2002 Jul 03  & 15 & 2.8 & 1.1 x 0.6 & -1  \\
  2002 Jul 19  & 15 & 3.1 & 1.1 x 0.6 & -5  \\
  2002 Aug 30  & 15 & 2.8 & 1.1 x 0.5 & -7  \\
  2002 Sep 20  & 15 & 3.3 & 1.1 x 0.5 & -5  \\
  2002 Oct 09  & 15 & 3.5 & 1.1 x 0.6 & -9  \\
  2002 Nov 15  & 15 & 3.2 & 1.1 x 0.5 & -8  \\
  2003 Jan 13  & 15 & 2.0 & 1.2 x 0.6 & -7  \\
  2003 Mar 05  & 15 & 1.6 & 1.0 x 0.5 &  2  \\
  2003 May 10  & 15 & 2.9 & 1.1 x 0.5 & -3  \\
  2003 May 26  & 15 & 2.2 & 1.0 x 0.5 & -1  \\
  2003 Jul 11  & 15 & 2.0 & 1.2 x 0.5 &  2  \\
  2003 Aug 28  & 15 & 2.0 & 1.0 x 0.5 & -5  \\
  2004 Apr 01  & 15 & 4.0 & 1.0 x 0.5 & -2  \\
  2004 Aug 09  & 15 & 3.8 & 1.0 x 0.6 & -5  \\
  2004 Sep 02  & 15 & 3.6 & 1.1 x 0.5 & -3  \\
  2004 Nov 05  & 15 & 2.8 & 1.0 x 0.6 & -5  \\
  2004 Dec 02  & 15 & 2.2 & 1.0 x 0.6 & -6  \\
  2005 Feb 14  & 15 & 2.2 & 1.1 x 0.5 & -1  \\
  2005 Mar 13  & 15 & 1.9 & 1.2 x 0.6 & -1  \\
  2005 May 14  & 15 & 1.9 & 1.4 x 0.5 & -11 \\
  2005 May 19  & 15 & 1.8 & 1.2 x 0.5 & -2  \\
  2005 May 21  & 15 & 2.0 & 1.2 x 0.5 &  2  \\
  2005 Jun 03  & 15 & 2.0 & 1.0 x 0.5 & -8  \\
  2005 Jul 14  & 15 & 2.9 & 1.2 x 0.5 & -2  \\
  2005 Aug 29  & 15 & 3.2 & 1.1 x 0.5 & -2  \\
  2005 Nov 14  & 15 & 3.3 & 1.2 x 0.5 & -13 \\
  2006 Apr 28  & 15 & 4.6 & 1.0 x 0.5 & -4  \\
  2006 May 12  & 15 & 4.1 & 1.2 x 0.6 & -4  \\
  2006 Sep 21  & 15 & 2.6 & 1.1 x 0.5 & -7  \\
  2006 Dec 06  & 15 & 2.6 & 1.2 x 0.5 & -12 \\
  2007 Jan 06  & 15 & 2.3 & 1.0 x 0.5 & -2  \\
  2007 Mar 02  & 15 & 1.9 & 1.0 x 0.6 & -2  \\
  2007 Apr 09  & 15 & 1.3 & 1.1 x 0.5 & -11 \\
  2007 Apr 18  & 15 & 1.4 & 1.0 x 0.6 & -7  \\
  2007 Jun 10  & 15 & 1.3 & 1.1 x 0.5 & -5 \\
  2007 Sep 06  & 15 & 1.7 & 1.0 x 0.6 &  5 \\
  2008 Jan 12  & 15 & 2.4 & 1.1 x 0.5 &  3 \\
  2008 May 01  & 15 & 2.8 & 1.0 x 0.5 & -6 \\
  2008 Jul 17  & 15 & 3.0 & 1.1 x 0.5 & -5 \\
  2008 Aug 06  & 15 & 3.0 & 1.1 x 0.5 & -7 \\ 
  2008 Sep 12  & 15 & 2.8 & 1.1 x 0.6 & -7 \\
  2008 Nov 19  & 15 & 2.5 & 1.1 x 0.5 & -4 \\
  2009 Feb 02  & 15 & 3.8 & 1.1 x 0.6 &  1 \\
  2009 Feb 05  & 15 & 3.9 & 1.2 x 0.5 & -10 \\
  2009 Mar 25  & 15 & 4.5 & 1.1 x 0.6 & -5 \\
  2009 May 24  & 15 & 4.3 & 1.0 x 0.6 & -4 \\
  2009 Aug 19  & 15 & 5.8 & 1.0 x 0.6 & -6 \\
  2009 Oct 25  & 15 & 6.5 & 1.1 x 0.5 & -9 \\
  2010 May 24  & 15 & 4.1 & 1.3 x 0.6 & -12 \\
  2010 Jul 09  & 15 & 4.3 & 1.1 x 0.5 & -11 \\
  2010 Aug 28  & 15 & 2.5 & 1.0 x 0.5 & -5 \\
  2010 Oct 18  & 15 & 4.5 & 1.0 x 0.5 & -5 \\
  2011 Feb 27  & 15 & 5.5 & 1.0 x 0.5 & -3 \\
  2011 Aug 26  & 15 & 5.0 & 1.1 x 0.6 & -7 \\
  2011 Dec 29  & 15 & 4.8 & 1.1 x 0.6 &  2 \\
  2012 may 24  & 15 & 4.7 & 1.0 x 0.5 &  0 \\
  2007 Jun 14  & 43 & 1.0 & 0.5 x 0.2 & -15 \\
  2007 Jul 12  & 43 & 0.9 & 0.4 x 0.2 & -8  \\
  2007 Aug 06  & 43 & 0.8 & 0.5 x 0.2 & -17 \\
  2007 Aug 30  & 43 & 1.2 & 0.4 x 0.2 & -9  \\
  2007 Sep 29  & 43 & 1.5 & 0.5 x 0.2 & -19  \\
  2007 Nov 01  & 43 & 2.0 & 0.5 x 0.2 & -21 \\
  2008 Jan 17  & 43 & 2.0 & 0.4 x 0.2 & -9  \\
  2008 Feb 29  & 43 & 1.4 & 0.4 x 0.2 & -16 \\
  2008 Jun 12  & 43 & 2.7 & 0.5 x 0.2 & -14 \\
  2008 Jul 06  & 43 & 2.2 & 0.4 x 0.2 & -11 \\ 
  2008 Aug 16  & 43 & 2.3 & 0.5 x 0.2 & -22 \\
  2008 Sep 10  & 43 & 2.1 & 0.5 x 0.2 & -16 \\
  2008 Nov 16  & 43 & 2.9 & 0.4 x 0.2 & -9  \\
  2008 Dec 21  & 43 & 2.7 & 0.4 x 0.2 & -7  \\
  2009 Jan 24  & 43 & 5.1 & 0.4 x 0.2 & -13 \\
  2009 Feb 22  & 43 & 4.7 & 0.4 x 0.2 & -15 \\
  2009 Apr 01  & 43 & 5.0 & 0.4 x 0.2 & -14 \\
  2009 May 30  & 43 & 3.5 & 0.4 x 0.2 & -7 \\
  2009 Jun 21  & 43 & 2.7 & 0.3 x 0.2 & -1 \\
  2009 Jul 26  & 43 & 3.7 & 0.3 x 0.2 & -1 \\
  2009 Aug 16  & 43 & 4.9 & 0.4 x 0.2 & -1 \\
  2009 Sep 16  & 43 & 5.1 & 0.4 x 0.2 &  1 \\
  2009 Oct 14  & 43 & 5.6 & 0.4 x 0.2 & -6 \\
  2009 Oct 16  & 43 & 7.4 & 0.5 x 0.2 & -21 \\
  2009 Oct 20  & 43 & 7.4 & 0.5 x 0.2 & -16 \\
  2009 Oct 25  & 43 & 5.4 & 0.4 x 0.2 & -5 \\
  2009 Nov 28  & 43 & 7.1 & 0.5 x 0.2 &  28 \\
  2010 Jan 10  & 43 & 8.2 & 0.4 x 0.2 & -4 \\
  2010 Feb 11  & 43 &3.3 & 0.4 x 0.2 & -7 \\
  2010 Mar 06  & 43 & 6.0 & 0.4 x 0.2 & -4 \\
  2010 May 19  & 43 & 1.8 & 0.4 x 0.2 & -11 \\
  2010 Jun 14  & 43 & 1.8 & 0.4 x 0.2 & -7 \\
  2010 Aug 01  & 43 & 1.4 & 0.3 x 0.2 & -7 \\
  2010 Aug 21  & 43 & 1.4 & 0.4 x 0.2 & -10 \\
  2010 Sep 18  & 43 & 3.6 & 0.4 x 0.2 & -16 \\
  2010 Oct 24  & 43 & .0 & 0.5 x 0.2 & -15 \\
  2010 Nov 01  & 43 & 3.4 & 0.5 x 0.2 & -4 \\
  2010 Nov 06  & 43 & 3.6 & 0.5 x 0.2 & -8 \\
  2010 Nov 13  & 43 & 2.8 & 0.4 x 0.2 & -5 \\
  2010 Dec 04  & 43 & 3.1 & 0.4 x 0.2 & -6 \\
  2011 Jan 02  & 43 & 3.4 & 0.5 x 0.2 & -17 \\
  2011 Feb 04  & 43 & 3.6 & 0.4 x 0.2 & -14 \\
  2011 Mar 01  & 43 & 3.7 & 0.4 x 0.2 & -12 \\
  2011 Apr 21  & 43 & 3.7 & 0.3 x 0.2 &  3  \\
  2011 May 22  & 43 & 4.9 & 0.4 x 0.2 & -16 \\
  2011 Jun 12  & 43 & 2.9 & 0.3 x 0.2 & -2 \\
  2011 Jul 21  & 43 & 2.7 & 0.4 x 0.2 & -6 \\
  2011 Aug 23  & 43 & 2.9 & 0.5 x 0.3 & -18 \\
  2011 Sep 16  & 43 & 2.1 & 0.4 x 0.2 & -9 \\
  2011 Oct 16  & 43 & 3.6 & 0.4 x 0.2 & -5 \\
  2011 Dec 02  & 43 & 2.7 & 0.4 x 0.2 & -8 \\
  2012 Jan 27  & 43 & 3.2 & 0.3 x 0.2 & -7 \\
  2012 Mar 05  & 43 & 3.4 & 0.4 x 0.2 & -10 \\
  2012 Apr 02  & 43 & 3.0 & 0.3 x 0.2 &  1 \\
  \enddata
  
  \end{deluxetable}


\begin{thebibliography}{}
\bibitem[Agudo et al. (2010)]{agu10} Agudo,I, Jorstad, S.G, Marscher, A.P., Larionov, V.M., G\'omez,J.L., Wiesemeyer, H.,  Thum, C., Gurwell,M., Heidt, J., \& D'Arcangelo, F.D., 2010, in Fermi Meets Jansky - AGN in
Radio and Gamma -Rays, ed. T. Savolainen et al. (Bonn: Max-Planck-Institute f\"ur Radioastronomie), 143
\bibitem[Agudo et al. (2012)]{agu12} Agudo, I., Marscher, A.P., Jorstad, S.G., G\'omez,J.L., Perucho, M.,
Piner, B.G., Rioja, M., \& Dodson, R. 2012, \apj, 747, 63

\bibitem[Bajkova (2001)]{baj01} Bajkova, A.T., 2001, Astron. Astrophys. Trans., 20, 393

\bibitem[H\"ogbom(1974)]{hog74} H\"ogbom, J.A. 1974, \aaps, 15, 417

\bibitem[Katz (1997)]{kat97} Katz, J.I., 1997, \apj, 478, 527

\bibitem[Kellermann et al. (2004)]{kel04} Kellermann, K.I., Lister, M.L., Homan, D.C., Vermeulen, R.C., Cohen, M.H., Ros, E., Kadler, M., Zensus, J.A., \& Kovalev, Y.Y., 2004, \apj, 609, 539

\bibitem[Kovalev et al. (2005)]{kov05} Kovalev, Y.y., Kellermann, K.I., Lister, M.L., Homan, D.C., Vermeulen, R.C., 
Cohen, M.H., Ros, E., Kadler, M., Lobanov, A.P., Zensus, J.A., Kardashev, N.S., Gurvits, L.L., Aller,H.D. 2005, \aj, 230, 2473-2505

\bibitem[Lister \& Homan (2005)]{lih05} Lister, M.L., \& Homan, D.C. 2005, \aj, 130, 1389

\bibitem[Reid (2006)]{rei06} Reid, Robert I. 2006 \mnras, 367, 1766, 1780

\bibitem[Rastorgueva et al.(2011)]{ras11} Rastorgueva, E.A., Wiik, K.J., Bajkova, A.T., Valtaoja, E., Takalo, L.O., 
Vetukhnovskaya, Y.N., \& Mahmud, M., 2001, \aap, 529,A2,12

\bibitem[Shephard (1997)]{she97} Shepherd,M.C. 1997, Astronomical Data Analysis Software and Systems VI, ASP Conf. Ser. 125, ed G. Hunt \& H.E Payne (San Francisco: ASP), 77 

\bibitem[Sillanp\"a\"a et al. (1988)]{sil88} Sillanp\"a\"a, A, Haarala, S., Valtonen, M.J., Sundelius, B., \& Byrd, G.G.  1988, \apj, 325, 628

\bibitem[Tateyama et al. (1999)]{tat99} Tateyama, C.E., Kingham, K.A., Kaufmann, P., Piner, B.G., Botti, L.C.L., \& 
De Lucena, A.M.P. 1999, \apj, 520, 627

\bibitem[Tateyama (2009)]{tat09} Tateyama, Claudio E., 2009, \apj, 705, 877

\bibitem [Tateyama and Kingham (2004)] {tak04} Tateyama, C.E., \& Kingham, K.A., 2004, \apj, 608, 149

\bibitem[Valtonen and Wiik (2012)]{vaw12} Valtonen, M.J., \& Wiik, K., 2012, \mnras, 421, 186

\bibitem [Villforth et al. (2010)] {vil10} Villforth, C., Nilsson, K., Heidt, J., Takalo, L.O., Pursimo, T., Berdyugin, A., Lindfors, E., Pasanen, M., Winiarski, M.,  \& Drozdz, M.  2010, \mnras, 402,2087


\end{thebibliography}
\end{document}